\documentclass[10pt,showpacs,twocolumn]{revtex4}
\usepackage{graphicx}
\usepackage{amsmath}
\usepackage{amssymb}
\usepackage{color}
\usepackage{}
\begin{document}

\preprint{APS/123-QED}

\title{Frequency shift in high order harmonic generation from isotopic molecules}

\author{Lixin He,$^{1}$ Pengfei Lan,$^{1,}$\footnote{pengfeilan@mail.hust.edu.cn} Chunyang Zhai,$^{1}$ Feng Wang,$^{1}$ Wenjing Shi,$^{1}$ Qingbin Zhang,$^{1}$ Xiaosong Zhu,$^{1}$ and Peixiang Lu$^{1,2,}$\footnote{lupeixiang@mail.hust.edu.cn}
}



\affiliation{%
 $^1$School of Physics and Wuhan National Laboratory for Optoelectronics, Huazhong University of Science and Technology, Wuhan 430074, China\\
 $^2$Laboratory of Optical Information Technology, Wuhan Institute of Technology, Wuhan 430205, China
}%

\date{\today}

\begin{abstract}
We report the first experimental observation of frequency shift in high order harmonic generation (HHG) from isotopic molecules H$_2$ and D$_2$.
It is found that harmonics generated from the isotopic molecules exhibit obvious spectral red shift with respect to those from Ar atom. The red shift is further demonstrated to arise from the laser-driven nuclear motion in isotopic molecules. By utilizing the red shift observed in experiment, we successfully retrieve the nuclear motion in H$_2$ and D$_2$, which agree well with the theoretical calculations from the time-dependent Schr\"{o}dinger equation (TDSE) with Non-Born-Oppenheimer approximation. Moreover, we demonstrate that the frequency shift can be manipulated by changing the laser chirp.

\end{abstract}                         
\maketitle

High order harmonic generation (HHG), a nonlinear non-perturbative process, occurs in the interaction of intense laser fields
with atomic or molecular gases \cite{hg1,hg2}. Nowadays, HHG has been of great interest to produce coherent attosecond pulses in the extreme
ultraviolet and soft x-ray regions \cite{at1,at2,at3,at4,at5,at6}, which provide an important tool for probing the ultrafast electronic
dynamics inside atoms and molecules \cite{dy1,dy2}. On the other hand, harmonics from molecules encode rich information of the molecular structures and dynamics, which has irritated a lot of investigations on high harmonic spectroscopy \cite{im1,im2,im3,im4,im5}.



Compared to HHG in atoms, molecular high order harmonic generation (MHHG) is more complex due to the additional degree of freedom of the nuclear motion.
Many works have been carried out to investigate the effects of nuclear motion on strong-field ionization \cite{tz,ts,gn,ot} and MHHG \cite{ad,adb,lf,xl,ha}.
Recently, 
by comparing the relative
intensities of MHHG from isotopic molecules (H$_2$ and D$_2$), the nuclear motion has been theoretically predicted by Lein \cite{lein} and experimentally detected by Baker \emph{et al.} \cite{baker}. Whereas, this method is restricted due to the propagation and some other inherent physical effects, such as two-center interference \cite{lein,sba}, which may affect the harmonic intensity.
On the other hand, it's well known that the isotopic effect leads to a frequency shift in the atomic spectrum \cite{whk}, which has been widely used to identify the species of the isotopes. The frequency shift here reflects the static structure of the isotopes. This stimulate us to ask whether the nuclear motion in intense laser fields will induce a frequency shift in the MHHG spectrum, and if so, can we retrieve the nuclear motion from the frequency shift in MHHG? However, the measurement of nuclear motion based on the frequency shift in MHHG from isotopic molecules is seldom investigated, except the recent theoretical discussion \cite{bian}.



In this letter, we report the first experimental observation of frequency shift in HHG from isotopic molecules H$_2$ and D$_2$. The results show that harmonics generated from isotopic molecules show obvious red shift with respect to those from Ar atom. The red shift is further demonstrated to originate from the nuclear motion in isotopic molecules. From the observed red shift, the nuclear vibrations of H$_2$ and D$_2$ are successfully retrieved, which agree well with the calculations from Non-Born-Oppenheimer time-dependent Schr\"{o}dinger equation (NBO-TDSE). In addition, we also demonstrate the method to manipulate the frequency shift of MHHG by a chirped laser field.

Our experiment is performed by using a commercial Ti:sapphire laser system (Legend Elite-Duo, Coherent, Inc.), which delivers the 30-fs, 800-nm pulses at a repetition rate of 1 kHz. The output laser pulse is focused to a 2-mm-long gas cell by a 600-mm focal-length lens. The stagnation pressure of the
gases is 30 torr and the gas cell is placed 2 mm after the laser focus to ensure the phase matching of the short quantum path.
The laser energy used in our experiment is about 1.5 mJ and the corresponding intensity is estimated to be $1.5\times10^{14}\ \mathrm{W/cm}^2$. The generated harmonic spectrum is detected by a home-made flat-field soft x-ray spectrometer.

\begin{figure}[htb]
\centerline{
\includegraphics[width=7.5cm]{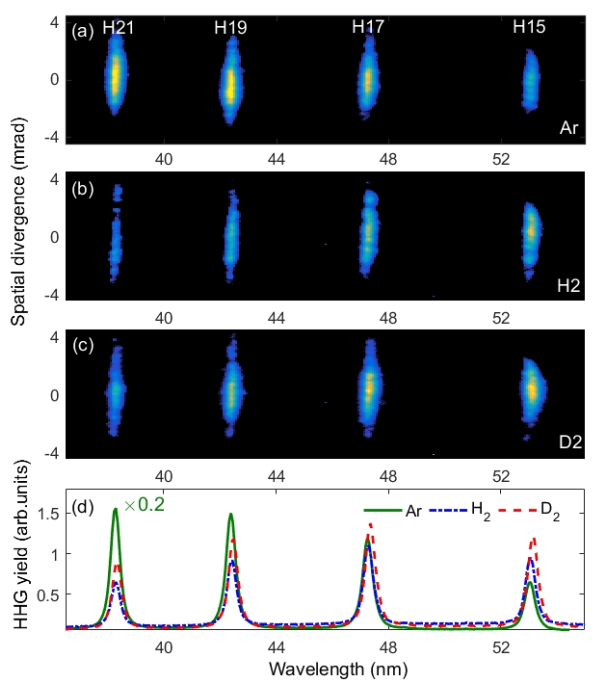}}
 \caption{\label{fig1}(Color online)  Measured harmonic spectra from (a) Ar, (b) H$_2$, and (c) D$_2$. (d) Spatially integrated HHG signals for the spectra in (a) (solid line), (b) (dash-dotted line) , and (c) (dashed line), respectively. For clarity, the solid line is multiplied by a factor of
0.2. The laser intensity is $1.5\times10^{14}\ \mathrm{W/cm}^2$ and the pulse duration is 30 fs.
 }
\end{figure}

Figures 1(a)-(c) display the spatially resolved harmonic spectra generated from atomic gas Ar and the hydrogen isotopes H$_2$ and D$_2$, respectively. The spatially integrated HHG signals are presented by the solid (Ar), dash-dotted (H$_2$), and dashed (D$_2$) lines in Fig. 1(d).
One can see that the harmonic intensities from D$_2$ are higher than those from H$_2$, which is in consistent with previous studies \cite{lein,baker,hm,kan}. More importantly, one can clearly see that the measured harmonics from H$_2$ and D$_2$ present obvious frequency shift with respect to those from Ar.
For clarity, we have normalized the intensities of harmonics 15-21 (H15-H21) from Ar (solid line), H$_2$ (dash-dotted line), and D$_2$ (dashed line).
As shown in Figs. 2(a)-(d), each harmonic from both H$_2$ and D$_2$ shows a red shift relative to that from Ar. While for D$_2$, the red shift is larger than that of H$_2$.



As is well known, HHG process in gas medium includes the individual response as well as the copropagation of laser and harmonic fields.
The propagation effect can possibly induce a frequency shift in HHG \cite{wmw,fb,chn}. However,
in our experiment the ionizations of the three gases are weak (below 4\%), and also the gas pressure is low. Then the frequency shift induced by the propagation effect will be inappreciable. For
further check, we have measured the harmonic spectra from the three gases with gas pressures changing from 15 torr to 35 torr. We find that the central wavelengths of each harmonic from the three gases are basically constant as the gas pressure increases (see Section A of the Supplementary material). This result suggests that the propagation effect on harmonic frequency shift can be ignored in our experiment.
Moreover, the experimental conditions used for HHG from H$_2$ and D$_2$ are exactly the same, the differences in the harmonic spectra can be mainly attributed to the individual response of isotope molecules in the driving laser field.

It has been reported that the nonadiabatic effect of the time-dependent laser intensity will induce a blue shift or red shift when HHG 
is dominant at the leading or falling part of the laser pulse \cite{kjs,mg,bian2}.
For H$_2$ (or D$_2$), the ionization rate depends sensitively on the internuclear distance $R$ \cite{tz,ts}.
Due to the laser-driven nuclear motion, the average internuclear distance at the falling part can be larger than that at the leading part of the laser pulse, which makes the ionization as well as the HHG signals stronger at the falling part, and therefore induces a strong red shift in the harmonic spectrum.
In contrast, since the laser intensity used in our experiment is far smaller than the ionization saturation threshold of Ar, the ionization and HHG of Ar atom mainly occurs at central part of the laser pulse and is symmetric with respect to the pulse center (t=0 a.u.). Then no obvious net shift exists in the harmonics from Ar. It can serve as a benchmark to evaluate the frequency shift of harmonics from the two isotopic molecules.
The frequency shift caused by the nonadiabatic effect can be given by
$\Delta\omega=\alpha_q\frac{\partial I(t)}{\partial t}$ \cite{mbg,psa,he}.
Here, $I(t)=I_0 \exp(\frac{-4ln(2) t^2}{\tau^2})$ with $I_0=1.5\times10^{14}\ \mathrm{W/cm}^2$ and $\tau=30$ fs
is the envelope of the laser pulse, and $\alpha_q$ is the phase coefficient of $q$th harmonic,
which can be evaluated according to the strong-field approximation (SFA) model \cite{3s1}.
Owing to the slower nuclear motion of heavier nuclei, the dominant harmonic emission of D$_2$ occurs later
than that of H$_2$. As a result, the HHG from D$_2$ experiences a more rapid change of the effective laser intensity (i.e., a larger value of $|\frac{\partial I(t)}{\partial t}|$),
which therefore gives rise to a larger red shift in the harmonic spectrum as observed in our experiment.
{Besides the nonadiabatic effect, the nuclear motion can lead to the variation of the ionization potential and the complex recombination dipole, which may affect the harmonic phase accumulated during the electron excursion and influence the MHHG \cite{le,fer}.
To evaluate these influences, we preform the simulation with the modified SFA model \cite{le},
which indicates that the frequency shift induced by these two effects is far smaller than our experimental observations (see Section B of the supplementary material).
Therefore, the main contribution to the frequency shift shown in Fig. 2 is attributed to the nuclear motion induced nonadiabatic effect \cite{bian}.}

For a quantitative view, we have presented the relative frequency shift of H15-H23 for H$_2$ (squares) and D$_2$ (circles) in Fig. 3(a).
The relative red shift is demonstrated to gradually decrease as the harmonic order increases.
{The experiment is also simulated by solving the non-Born-Oppenheimer time dependent Schr\"{o}dinger equation (NBO-TDSE) (see Section C of the Supplementary material). The calculated frequency shifts of HHG from H$_2$ and D$_2$ are presented by the solid lines in Fig. 3(a). One can see that the experimental observations are in agreement with the theoretical simulations.
Some difference in quantity may arise from the uncertainties of experimental parameters.}
Based on the measured red shift, we can estimate the average time where the dominant harmonic emission occurs for each harmonic in terms of $\Delta\omega=\alpha_q\frac{\partial I(t)}{\partial t}|_{t=t_{ave}}$.
The results are presented in Fig. 3(b). Obviously, the dominant harmonic emission of D$_2$ (circles) occurs even later than that of H$_2$ (squares) 
by about 200 as.

\begin{figure}[htb]
\centerline{
\includegraphics[width=8.5cm]{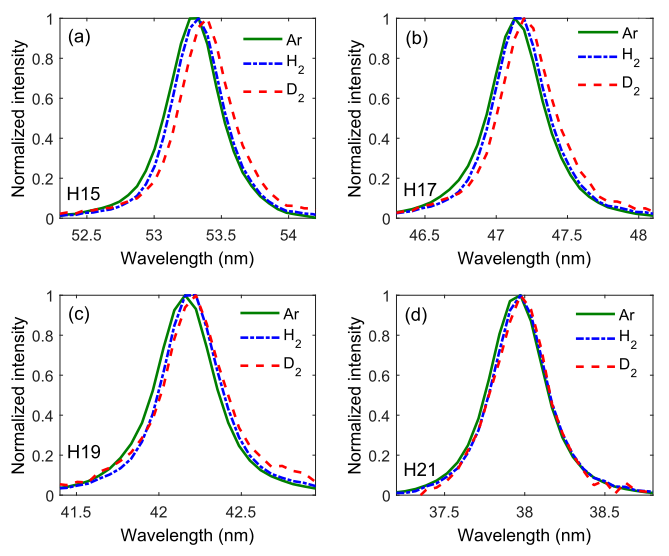}}
 \caption{\label{fig2}(Color online) (a)-(d) Normalized harmonic signals of H15-H21 for Ar (solid line), H$_2$ (dash-dotted line), and D$_2$ (dashed line), respectively.
 }
\end{figure}

Next, we consider to retrieve the nuclear motion from the observed frequency shift in MHHG.
As mentioned above, the frequency shift mainly arises from the asymmetry of the ionization (and so HHG) with respect to the center of laser pulse (t=0 a.u.) \cite{bian}. Previous studies have shown that the ionization rate of H$_2$ (D$_2$) is approximately linearly dependent on internuclear distance $R$ before it reaches 2 a.u. \cite{as}. Thus the relative frequency shift of $q$th harmonic with respect to $q\omega_0$ can be estimated as
\begin{eqnarray}
\frac{\Delta\omega}{\omega_0}=\frac{ \sum\limits_{t_i<0} R(t_i)-\sum\limits_{t_i>0} R(t_i)} {\sum\limits_{t_i} R(t_i)},
\end{eqnarray}
where $\omega_0$ is the frequency of the driving laser, $t_i$ is the ionization moment of the electron (contributing to $q$th harmonic generation) in each half optical cycle. $t_i<0$ and $t_i>0$ mean the ionization occurs at the leading and falling edges of laser pulse, respectively. For a given harmonic, $t_i$ can be calculated according to the three-step model \cite{3s1,3s2}.
To retrieve the nuclear motion, we consider to employ the simple linear harmonic oscillator model \cite{lad} to describe the nuclear motions of H$_2$ and D$_2$.
\begin{figure}[htb]
\centerline{
\includegraphics[width=8.5cm]{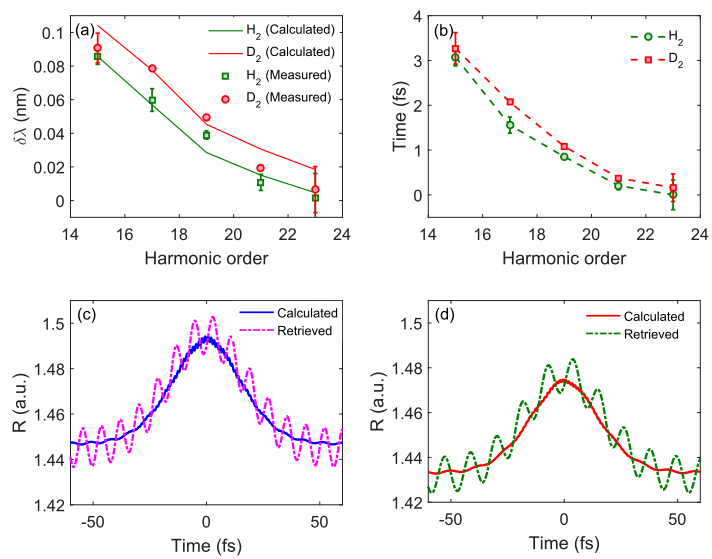}}
 \caption{\label{fig3}(Color online) (a) Measured and calculated frequency shift $\delta\lambda$ in MHHG with respect to harmonics from Ar and (b) the retrieved average emission time of MHHG as a function of the harmonic order. Squares and circles are for H$_2$ and D$_2$, respectively. (c) Calculated (solid line) and experimentally retrieved (dash-dotted line) nuclear vibration of H$_2$. (d) Same to (c), but for D$_2$.
 }
\end{figure}
Since the ionization and dissociation probabilities in our experimental condition are very low, the harmonic oscillator model can work well.
In the harmonic oscillator model,
the potential $V(r)$ of H$_2$ (or D$_2$) can be approximatively expressed as $V(r)=V_0+\frac{k}{2}(r-R_e)^2$ (a typical potential of linear harmonic oscillator), where $V_0$ and $k$ are constants and $R_e$ is the equilibrium distance ($\sim$1.4 a.u.) of H$_2$ and D$_2$.
Then the laser-driven nuclear motion of H$_2$ and D$_2$ can be derived in the form of (for details, see Section D of the Supplementary material)
\begin{eqnarray}
R(t)=A\sin(\Omega t+\phi)+B I(t)+R_e.
\end{eqnarray}
On the right side of Eq. (2), the first term denotes the inherent harmonic vibration, $A$, $\Omega$, and $\phi$ are the corresponding amplitude, frequency, and phase of the vibration. The second term represents the laser-nucleus interaction.
Inserting Eq. (2) into Eq. (1), the frequency shift of a specific harmonic can be expressed as a function of $A, B$, $\Omega$, and $\phi$. {By fitting the observed frequency shifts of H15-H23 to Eq. (1) with the least square method, the four parameters can be determined.} Then the nuclear motion $R(t)$ can be retrieved. Figures 3(c)-(d) show the retrieved nuclear vibrations (dash-dotted line) of H$_2$ and D$_2$, respectively.
The results calculated from the NBO-TDSE \cite{ad,tkr,ss} are also presented as the solid lines in Figs. 3(c)-(d) (for calculation details, see Section D of the Supplementary material).
{One can see that the main structures of the retrieved nuclear motion $R(t)$, such as the dynamic range and the overall trend,
agree well with the theoretical predictions.
But the retrieved nuclear motion shows much deeper modulation, which is due to the inherent harmonic vibration of the harmonic oscillator model.}

\begin{figure}[htb]
\centerline{
\includegraphics[width=8.5cm]{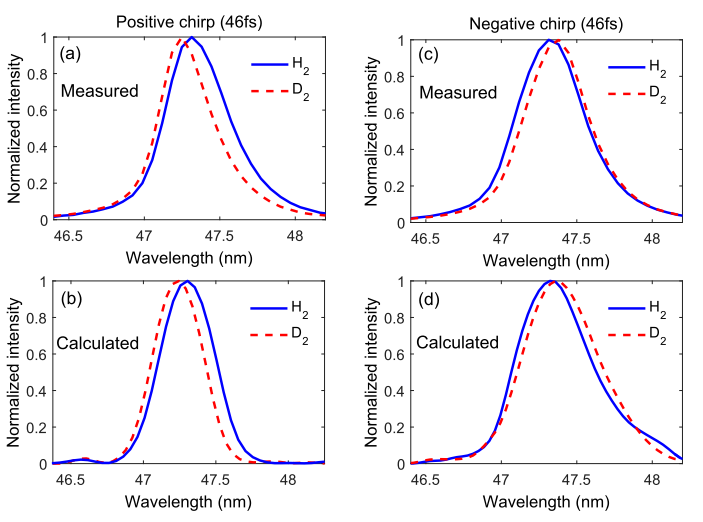}}
 \caption{\label{fig4}(Color online)  (a) Measured and (b) calculated harmonic signals of H17 in the laser field with a positive chirp.
(c)-(d) Same to (a)-(b), but for the field with a negative chirp. The solid and dashed lines are for H$_2$ and D$_2$, respectively. The laser intensity here is $1.5\times10^{14}\ \mathrm{W/cm}^2$.
 }
\end{figure}

On the other hand, we consider to manipulate the frequency shift in MHHG by using a chirped laser field.
In our experiment, we adjust the distance between the gratings in the compressor of the driving laser to introduce a linear chirp. The duration of the chirped laser pulse is measured by a autocorrelator.
Figure 4(a) shows the measured H17 of H$_2$ (solid line) and D$_2$ (dashed line) in a positively chirped laser field (the laser duration is about 46 fs). In this case, harmonic emission from D$_2$ exhibits a blue shift with respect to that from H$_2$. This is opposite to the result observed in the chirp-free field [Fig. 2(b)]. While in a negatively chirped field [Fig. 4(c)], the red shift of harmonic emission from D$_2$ with respect to that from H$_2$ still holds.
We have also calculated harmonic generation
in the chirped laser pulses by using the experimental parameters. Corresponding results are shown in Figs. 4(b) and 4(d). One can see that the experimental observations are consistent with the theoretical calculations.

\begin{figure}[htb]
\centerline{
\includegraphics[width=8.5cm]{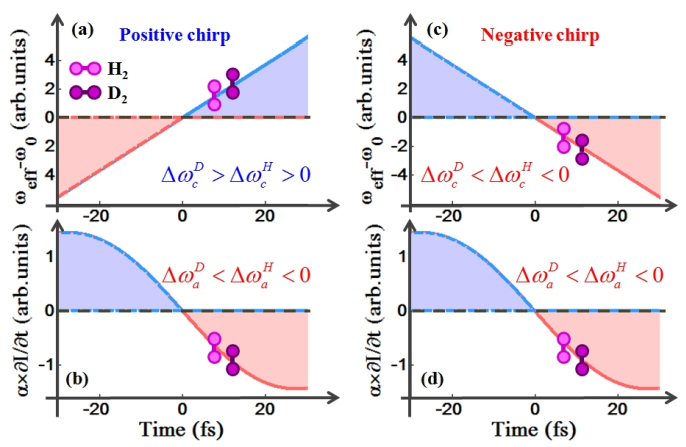}}
 \caption{\label{fig5}(Color online) Sketch of the harmonic frequency shift in a chirped laser field. (a)-(b) show the effects of laser chirp and the time-dependent laser intensity on the harmonic frequency shift in a positively chirped laser field. (c)-(d) are same to (a)-(b), but in a negatively chirped laser field.
 }
\end{figure}

In the chirped laser field, not only the time-dependent laser intensity but also the laser chirp can affect the frequency shift of MHHG.
The total frequency shift can be given by $\Delta\omega=\Delta\omega_a+\Delta\omega_c$, where $\Delta\omega_a$ and $\Delta\omega_c$ represent the frequency shift induced by the time-dependent laser intensity and laser chirp, respectively.
In Figs. 5(b) and (d), we show the values of $\alpha_q \frac{\partial I(t)}{\partial t}$ (frequency shift due to the time-dependent laser intensity) as a function of time in a positively and negatively chirped laser field, respectively.
As mentioned above, the MHHG is dominant at the falling part of the laser pulse due to the nuclear motion.
Therefore, for the laser field with either positive or negative chirps, the time-dependent laser intensity will introduce a red shift in MHHG, and the red shift of D$_2$ is larger than that of H$_2$, i.e., $\Delta\omega_a^{D}<\Delta\omega_a^{H}<0$ [see  Figs. 5(b) and (d)].
While in a positively chirped field, the effective laser frequency $\omega_{eff}=\omega_0+\beta t$ [$\beta>0$($<0$) for the positively (negatively) chirped pulse] at the falling edge is larger than the central frequency $\omega_0$ (at t=0 a.u.) [see Fig. 5(a)], which therefore induces a blue shift in MHHG. Since the dominant HHG from D$_2$ occurs later than that from H$_2$, the effective laser frequency for D$_2$ is larger that that for H$_2$. As a result, HHG from D$_2$ shows a larger blue shift than that from H$_2$, i.e., $\Delta\omega_c^{D}>\Delta\omega_c^{H}>0$. In this case,
the blue shift induced by the positive chirp can compensate the red shift induced by the time-dependent laser intensity. By changing the laser intensity or laser chirp, either blue or red shift can be observed in the MHHG.
On the contrary, in a negatively chirped field, the negative chirp introduces a red shift in MHHG, and the red shift of D$_2$ is also larger than that of H$_2$, i.e., $\Delta\omega_c^{D}<\Delta\omega_c^{H}<0$ [see Fig. 5(c)]. The negative chirp will enlarge the red shift induced by the time-dependent laser intensity. Therefore, in a negatively chirped field, one can only see the red shift in MHHG, and HHG from D$_2$ always presents a red shift with respect to that from H$_2$.

In summary, we have experimentally investigated HHG from isotopic molecules H$_2$ and D$_2$. The observed harmonic spectra of H$_2$ and D$_2$ show obvious red shift as compared to that of Ar atom. This phenomenon is primarily attributed to the laser-driven nuclear motion in H$_2$ and D$_2$, which strengthens the ionization and harmonic emission at the falling part of the laser pulse. By using a linear harmonic oscillator model, we successfully retrieve the nuclear vibrations of H$_2$ and D$_2$ from the observed frequency shift. Moreover, we show that the frequency shift in MHHG can be effectively manipulated by using a chirped field.
Under suitable laser conditions (laser intensity, laser chirp), the spectral red shift can be changed to a blue shift in the positively chirped pulse. While in a negatively chirped pulse, only the red shift can be observed.

In previous studies by Lein \cite{lein} and Baker \emph{et al.} \cite{baker},
the intensity ratios of HHG from isotopic molecules
reveal the nuclear dynamics of H$_2^+$ and D$_2^+$ within the time window from ionization to recombination in one laser cycle. While in the present work, the observed frequency shift provides a monitoring of the nuclear vibrations of H$_2$ and D$_2$ at each ionization moment in the laser pulse. Our work reveals a different physical process and is complementary with these previous studies \cite{lein,baker} in probing nuclear dynamics.

We gratefully acknowledge X. B. Bian, A. D. Bandrauk and T. Ozaki for
valuable discussions. This work was supported by the National Natural Science Foundation of China under Grants No. 11234004, 61275126, 11422435 and 11404123. Numerical simulations presented in this paper were
carried out using the High Performance Computing experimental testbed in SCTS/CGCL (see http://grid.hust.edu.cn/hpcc).

\end{document}